# Strong Dzyaloshinskii-Moriya Interaction in Monolayer CrI$_3$ on Metal Substrates

Fan Zhang[1], Xueao Li[1], Yabei Wu[2,3], Xiaolong Wang[1], Jijun Zhao[1*], Weiwei Gao[1*]

1. Key Laboratory of Material Modification by Laser, Ion and Electron Beams (Dalian University of Technology), Ministry of Education, Dalian 116024, China

2. Department of Materials Science and Engineering and Shenzhen Institute for Quantum Science & Engineering, Southern University of Science and Technology, Shenzhen, Guangdong, 518055, China

3. Guangdong Provincial Key Lab for Computational Science and Materials Design, and Shenzhen Municipal Key Lab for Advanced Quantum Materials and Devices, Southern University of Science and Technology, Shenzhen, Guangdong 518055, China

**ABSTRACT**

Dzyaloshinskii-Moriya interaction (DMI) is the primary mechanism for realizing real-space chiral spin textures, which are regarded as key components for the next-generation spintronics. However, DMI arises from a perturbation term of the spin-orbit interaction and is usually weak in pristine magnetic semiconductors. To date, large DMI and the resulting skyrmions are only realized in a few materials under stringent conditions. Using first-principles calculations, we demonstrate that significant DMI occurs between nearest-neighbor Cr atoms in two-dimensional (2D) magnetic semiconductor CrI$_3$ on Au or Cu substrates. This exceptionally strong DMI is generated by the interfacial charge transfer and weak chemical interactions between chromium halides and metal substrates, which break the spatial inversion symmetry. These findings highlight the significance of substrate effects in 2D magnets and expand the inventory of feasible materials with strong DMI.

* Corresponding Authors. Email: weiweigao@dlut.edu.cn, zhaojj@dlut.edu.cn

# INTRODUCTION

The applications of new magnetic systems and phenomena in electronics and spintronics have transformed and will continue to influence the way in which human process and store information [1–3]. Under the current technological framework, a slowdown in Moore's law [4,5] is likely when the lithography process approaches the sub-nanometer limit soon. Meanwhile, scientists are actively looking for new information carriers, lowering energy costs, and introducing new degrees of freedom in the information manipulation processes. For example, topologically protected spin textures such as skyrmions are important candidates as information carriers [6–10]. So far, several mechanisms have been proved to induce skyrmions, such as long-ranged magnetic dipolar interaction [11,12], the Dzyaloshinskii-Moriya interaction (DMI) [13,14], frustrated exchange interaction [15], and four-spin exchange interaction [16]. Among these mechanisms, DMI is arguably the most widely used one for stabilizing chiral spin textures.

According to Moriya's theory [14], DMI requires three key ingredients: spin-orbit coupling (SOC), magnetic exchange, and inversion asymmetry. It belongs to super-exchange interaction effect that originates from a second order perturbation term proportional to the strength of spin-orbit interaction and is usually much weaker than typical Heisenberg exchange interactions. DMI has been observed in bulk non-centrosymmetric materials such as $BiFeO_3$ and MnSi [6,18–20]. Weak DMI can also appear in centrosymmetric systems, such as $Fe_2O_3$ and $LaMnO_3$ [13,17], where spatial inversion symmetry is broken locally. Compared to single-phase semiconductors, DMI is known to be significantly stronger at metallic interfaces. For instance, Fe/Ir(111) [16] and some heterojunctions of ferromagnetic thin films with heavy metals [21–23] have been reported to host strong DMI and skyrmions.

Two-dimensional (2D) magnetic materials have recently emerged as ideal candidates for creating nanoscale spintronic and electronic devices. The majority of 2D van der Waals (vdW) magnetic materials, however, have the spatial inversion between the nearest-neighbor spin sites, which intrinsically prohibit strong DMI between them.

Several recent theoretical studies proposed that Janus materials [24,25] and intrinsic 2D multiferroics [26,27] with inversion asymmetry and heavy elements can potentially produce DMI that is strong enough to stabilize magnetic spin textures. To our knowledge, only a few Janus materials, such as MoSSe, have been successfully synthesized and they usually require well-crafted experimental procedures [28]. Additionally, it was proposed to generate DMI interaction in a centrosymmetric magnetic material like $CrI_3$ by applying a strong out-of-plane electric field [29], or by constructing the multiferroic vdW heterostructure [30,31]. In experiments, researchers found $Fe_3GeTe_2$, a vdW layered metal with itinerant ferromagnetism [32], can have strong DMI and host skyrmions when it is assembled into heterostructures with other materials [33,34]. Additionally, large DMI causes room-temperature skyrmion lattice in 50% Co-doped $Fe_5GeTe_2$ thin film [35]. Nevertheless, candidate materials (especially semiconductors) showing large DMI are highly limited. It is therefore imperative to discover new materials, preferably belonging to 2D vdW materials, which can be easily realized in experiments and can produce strong DMI.

Nowadays, constructing heterostructures and using substrates to induce emergent phenomena or tunning physical properties of 2D materials are efficient strategies of material engineering. For example, substrates are known to play key roles in influencing the electronic and optical properties of 2D materials, including exciton binding energies, quasiparticle band gaps, ionization potentials, and so on [36–41]. The impacts of substrates on the magnetic properties of 2D van der Waals materials deserve in-depth investigations [40,42]. Using density functional theory (DFT) calculations, herein we comprehensively study the effect of substrates such as hexagonal BN (h-BN), graphene, Au and Cu on prototypical 2D magnetic semiconductors $CrX_3$ (X = Cl, I), with a focus on DMI strength. Notably, even in centrosymmetric semiconductors like $CrI_3$ and $CrCl_3$, we find significant DMI when they form heterostructures with metal substrates like Au, and Cu. Both interfacial charge transfer and structural distortion contribute to the broken inversion symmetry, resulting in strong DMI.

**COMPUTATION METHODS**

DFT calculations were carried out using plane-wave and projector augmented wave approach as implemented by Vienna ab initio simulation package (VASP 5.4.4) [43]. We used the Perdew-Burke-Ernzerhof (PBE) functional [44] and the DFT-D3 correction [45] to treat exchange-correlation and Van der Waals interaction, respectively. The configurations of valence electrons considered in PAW potentials for Cu, Au, Cr, and I are $3d^{10}4s^1$, $5d^{10}6s^1$, $3d^54s^1$, and $5s^25p^5$, respectively. A plane-wave cutoff of 400 eV, a 2×2×1 Monkhorst−Pack k-point mesh for $CrI_3$/BL h-BN, $CrI_3$/graphene, $CrCl_3$/Au(111) and $CrI_3$/Au(111), and a 3×3×1 k-point mesh for $CrI_3$/Cu(111) were used [46]. To reduce the artificial interactions between periodic images, a 20 Å vacuum space perpendicular to the slab model was added. To account for the correlation effects from 3d electrons of Cr in a more faithful manner, we used GGA + $U$ [47,48] with Hubbard $U_{eff} = U - J = 2.6$ eV, which is consistent with several previous calculations [49,50]. All structures were optimized until the energy differences between consecutive ionic steps are less than $10^{-4}$ eV/supercell and the force felt by each atom is less than 0.01 eV/Å. Structural relaxations were performed without full-relativistic effects. All the metal substrates are modeled with three atomic layers. We checked the calculation results with metal substrates of four atomic layers, which do not change our conclusions (see Supplemental Material (Table S1) for more details [51]).

After structural optimization, we performed calculations with full-relativistic effects and non-collinear magnetism to extract the magnetic exchange parameters. The magnetic exchange parameters between nearest-neighbor Cr pairs were calculated using the four-state energy mapping method [52] with the supercells shown in Fig. 1. The single-ion MAE was calculated using the following equation:

$$MAE = E(\parallel) - E(\perp)$$

where $E(\parallel)$ denotes the energies calculated with Cr magnetic moments lying within the plane of monolayer $CrX_3$, and $E(\perp)$ stands for those with Cr magnetic moments pointing out-of-plane. MAE > 0 (< 0) corresponds to out-of-plane (in-plane) magnetic anisotropy. Constraints of magnetic moment directions were applied to ensure the system converge to the required magnetic configuration. To obtain accurate magnetic parameters, self-consistent calculations of noncolinear magnetic configuration were converged within $10^{-2}$ meV/supercell.

The binding energy between CrX$_3$ (X = Cl, I) and substrate is defined by the following equation:

$$E_{\rm b} = E_{\rm CrX_3/substrate} - E_{\rm CrX_3} - E_{\rm substrate}$$

where $E_{\rm CrX_3/substrate}$ is the total energy per area of the CrX$_3$/substrate heterostructures, $E_{\rm CrX_3}$ and $E_{\rm substrate}$ represent the total energy of freestanding monolayer CrX$_3$ and freestanding substrates, respectively. $E_{\rm CrX_3}$ and $E_{\rm substrate}$ were calculated with fixed lattice parameters the same as the supercell of corresponding heterostructure.

## RESULTS AND DISCUSSIONS

We have investigated five "CrX$_3$/substrate" (X = Cl, I) heterostructures to unveil the role of substrate on tunning magnetic exchange parameters of CrX$_3$. The substrates considered here include a polar insulator (bilayer hexagonal boron nitride with AB-stacking, abbreviated as BL h-BN thereafter) with an out-of-plane dipole field, a semimetal (graphene), and two metals (gold and copper). For metal substrates, the (111) surface of gold (Au) and copper (Cu) are chosen since they are stable and widely adopted as substrates for 2D materials like MoS$_2$ and graphene in laboratory [53–55]. The top and side views of the supercells for five heterostructures are shown in Fig. 1 [56,57]. These supercells are constructed such that the lattice mismatch between 2D magnets and substrates are small enough, i.e., from 0.8% to 2.9%. For example, the supercell of CrI$_3$/Au(111) heterostructure is constructed by combining a $\sqrt{3} \times \sqrt{3}$ supercell of monolayer CrI$_3$ with a 4×4 supercell of Au(111) surface in three-atomic-layer slab model, resulting in a lattice mismatch of 1.4%. The lattice mismatch and the optimized interlayer distance between 2D magnets and substrates are summarized in

Table 1.

The spin-dependent part of the ground-state electronic energy is mapped to the following classical spin Hamiltonian:

$$H_{spin} = \sum_{\alpha}^{x,y,z} \sum_{i<j} J_{ij}^{\alpha\alpha} S_i^\alpha S_j^\alpha + \sum_{i<j} \vec{D}_{ij} \cdot (\vec{S}_i \times \vec{S}_j) + A_z \sum_i (S_i^z)^2 \quad (1)$$

where $\vec{S}_i = [S_i^x, S_i^y, S_i^z]$, $J_{ij}^{\alpha\beta}$ stands for the exchange coupling between site $i$ and site $j$, $\vec{D}_{ij} = [D_{ij}^x, D_{ij}^y, D_{ij}^z]$ is the DMI vector, and $A_z$ describes the single-ion magnetic anisotropy. The exchange coupling parameters $J_{ij}^{\alpha\beta}$ can be extracted using a four-state energy mapping method [52] combined with full-relativistic DFT calculations (see Method Section for more details). In the following discussion, we consider the isotropic part of the Heisenberg exchange interaction to be $J_{ij} = \frac{1}{3}(J_{ij}^{xx} + J_{ij}^{yy} + J_{ij}^{zz})$. The elements of DMI vectors are related to the off-diagonal exchange parameters $J_{ij}^{\alpha\beta}$ ($\alpha \neq \beta$) by the following equations:

$$D_{ij}^x = \frac{1}{2}(J_{ij}^{yz} - J_{ij}^{zy}), \quad D_{ij}^y = \frac{1}{2}(J_{ij}^{zx} - J_{ij}^{xz}), \quad D_{ij}^z = \frac{1}{2}(J_{ij}^{xy} - J_{ij}^{yx}) \quad (2)$$

We are interested in the ratio between the magnitude of $|J_{ij}|$ and that of DMI vector $D_{ij} = \sqrt{(D_{ij}^x)^2 + (D_{ij}^y)^2 + (D_{ij}^z)^2}$. A larger $\frac{D_{ij}}{|J_{ij}|}$ suggests a stronger tendency to form non-collinear magnetic ordering. Earlier studies show that a ratio $\frac{D_{ij}}{|J_{ij}|}$ of over 10% [24,25,58] indicates the possibility of stable chiral spin textures.

We calculated magnetic exchange parameters $J_{ij}$ and $\vec{D}_{ij}$ between all nearest-neighbor Cr pairs. The magnitudes of averaged exchange parameters $|J|$ and $D$ in five "CrX$_3$-substrate" heterostructures are presented in Fig. 2(a). Clearly, the magnetic exchange parameters highly depend on substrates. The sizes of DMI in CrI$_3$/Au(111) and CrI$_3$/Cu(111) are significantly larger than the other three heterostructures. The ratios $\frac{D}{|J|}$ are compared in Fig. 2(b). Remarkably, the $\frac{D}{|J|}$ ratio is about 20% in CrI$_3$/Au(111) and CrI$_3$/Cu(111) systems. Such a large ratio between DMI and Heisenberg exchange interaction is even larger than those of prototypical materials hosting magnetic skyrmions under suitable conditions, such as MnSi [59],

Cu$_2$OSeO$_3$ [60], and Rh/Fe/Ir [61]. Overall, Fig. 2 demonstrates a positive correlation between the metallicity of substrate and the strength of induced DMI. Comparing Fig. 2 and Table 1, one can also see that stronger binding energy between the substrate and CrX$_3$ tends to enhance DMI.

To gain a detailed understanding of the effects of substrate on the magnetic properties of CrX$_3$, we performed further analysis on each heterostructure. We start from the system consisting of monolayer CrI$_3$ deposited on BL h-BN (see Fig. 1(a)). The BL h-BN we investigated here has AB-stacking order, in which the boron atoms of the top layer align with the center of hexagonal rings of the bottom layer [62]. The BL h-BN in AB-stacking has been demonstrated to be an out-of-plane ferroelectric in a recent experiment [62]. Together, the ferroelectric BL h-BN substrate and ferromagnetic monolayer CrI$_3$ form a multiferroic van der Waals heterostructure. The magnetic exchange parameters of selected nearest-neighbor Cr pairs in CrI$_3$/BL h-BN are summarized in Supplemental Material (Table S2) [51]. We find the average of Heisenberg exchange interaction $J$ is $-4.9$ meV, which is close to the exchange parameter $J$ of freestanding CrI$_3$ monolayer. Indeed, our calculated $J = -4.1$ meV for freestanding CrI$_3$ agrees well with an earlier study which obtained $J = -4.5$ meV [63]. The averaged magnitude of DMI between nearest-neighbor Cr pairs is about 0.046 meV, leading to a small ratio $\frac{D}{|J|} \approx 0.9\%$. Such a small ratio in CrI$_3$/BL h-BN heterostructure suggests the BL h-BN substrate in AB-stacking has a weak effect on generating DMI in monolayer CrI$_3$. When the single-layer CrI$_3$ combines with graphene to form a heterostructure (as shown in Fig. 1(b)), the average exchange parameters $|J|$ and $D$ change to 4.7 meV and 0.055 meV, respectively. Compared to CrI$_3$/BL h-BN, the ratio $\frac{D}{|J|}$ of CrI$_3$/graphene increases to 1.2%, accompanied by a boost of binding energy between CrI$_3$ and graphene, which will be discussed later.

To prove our guess that the free electrons from substrates can enhance DMI, we investigated the effects of metals, such as Au and Cu, as substrates for monolayer chromium halides. The exchange parameters of all the nearest-neighbor Cr pairs within the supercells of CrCl$_3$/Au(111), CrI$_3$/Au(111), and CrI$_3$/Cu(111) are calculated and summarized in Supplemental Material (Table S4-S6) [51]. Interestingly, the exchange parameters $\vec{D}_{ij}$ and $J_{ij}$ between Cr pairs show large fluctuations for CrCl$_3$ and CrI$_3$ deposited on metal substrates. Notably, the Heisenberg exchange $J_{ij}$ even changes sign between specific Cr pairs, such as Cr2-Cr3 pair in CrCl$_3$/Au(111) shown in Fig. 1(c). Moreover, for some Cr pairs, e.g., Cr2-Cr6 in CrI$_3$/Cu(111), the magnitude of DMI vector is even larger than that of Heisenberg exchange interaction. Overall, Au and Cu substrates significantly enhance the DMI between nearest-neighbor Cr pairs. For example, CrCl$_3$/Au(111) exhibits a non-negligible DMI that leads to $\frac{D}{|J|} = 2.6\%$. And CrI$_3$/Au(111) and CrI$_3$/Cu(111) possess a remarkably high ratio $\frac{D}{|J|}$ of 20%, that is approximately 20 times of that in CrI$_3$/BL h-BN and CrI$_3$/graphene. The significant $\frac{D}{|J|}$ value of CrI$_3$/Au(111) and CrI$_3$/Cu(111) suggests that combining 2D magnetic semiconductors and metal substrates is an effective way to induce considerable DMI. These heterostructures can be readily realized by existing experimental techniques, without resorting to strong electric field or synthesis of Janus materials.

The strong DMI and large fluctuation of exchange parameters induced by metal substrates can be explained by two main factors, namely, the interfacial charge transfer and weak bonds formed between the magnetic monolayer and substrates. These two factors cooperatively create asymmetrical charge redistributions in the top- and bottom-layer iodine atoms of CrI$_3$. To visualize the redistribution of electron charge density $\Delta\rho(\vec{r})$ caused by substrate, we compute

$$\Delta\rho(\vec{r}) = \rho_{\text{CrX}_3/\text{Substrate}}(\vec{r}) - \rho_{\text{Substrate}}(\vec{r}) - \rho_{\text{CrX}_3}(\vec{r}) \quad (3)$$

where $\rho_{\text{Substrate}}$, $\rho_{\text{CrX}_3}$, and $\rho_{\text{CrX}_3/\text{Substrate}}$ are the charge densities of the substrate, freestanding monolayer CrX$_3$, and CrX$_3$/substrate heterostructures, respectively. Fig. 3(a) and 3(b) show $\Delta\rho(\vec{r})$ within the planes of top-layer and bottom-layer iodine atoms in CrI$_3$/BL h-BN and CrI$_3$/Cu(111). Due to the differences in electrostatic potential between the top and bottom iodine atoms layers, the charge density redistribution $\Delta\rho(\vec{r})$ shows asymmetric features in both heterostructure systems. For example, the charge density redistribution around the top-layer iodine atoms induced by bilayer h-BN is negligible, while that within the bottom layer of iodine atoms is on the order $10^{-4}$ e/Bohr$^3$. The magnitude of $\Delta\rho(\vec{r})$ induced by BL h-BN is over one-order-of-magnitude smaller than that induced by Cu(111) substrate, because the insulating nature of h-BN hinders the transfer of sufficient electrons to CrX$_3$. Additionally, compared to CrI$_3$/Cu(111), the larger interlayer spacing between BL h-BN and CrI$_3$ also weakens the interlayer interactions. As a result, the out-of-plane dipole field from BL h-BN is too weak to induce a sizable redistribution of the charge density required for creating large DMI in monolayer CrI$_3$.

The large fluctuation of magnetic exchange parameters is a consequence of structural distortions caused by the weak bonds between metal substrates and the halogen ions X of monolayer CrX$_3$. Table 1 also presents the binding energy to identify the interaction strength between CrX$_3$ and substrates. The binding energy of CrX$_3$ with metal substrate is much higher than that of insulator and semimetal. In particular, the binding energy between CrI$_3$ and Au(111) or Cu(111) is over 50 meV/Å$^2$, which is much stronger than van der Waals interactions [64] but still much weaker than typical covalent bonds [65]. For reference, the exfoliation energies of graphene and phosphorene are 32 meV/Å$^2$ and 23 meV/Å$^2$, respectively [66]. The weak bonds between metal substrate and CrI$_3$ slightly distort the atoms from high-symmetry positions and break the symmetry that is present in freestanding CrI$_3$. In Supplemental Material (Fig. S1) [51], we plot the displacements of CrI$_3$ deposited on Au(111) and Cu(111) substrates, respectively. Both Au(111) and Cu(111) substrates induce non-negligible displacements of I and Cr atoms. Due to the distortions brought by the interlayer interactions, the $\Delta\rho(\vec{r})$ of CrI$_3$/Cu(111) lacks the C$_3$ symmetry, as shown in Fig. 3(b). Interestingly, the binding energy between CrCl$_3$ and Au(111) substrate is only 30 meV/Å$^2$, which is comparable to the exfoliation energy of phosphorene and suggests the interaction between CrCl$_3$ and Au(111) is mainly of vdW type. Based on our analysis, we speculate a heterostructure between Van der Waals magnets and metal substrates with small lattice mismatch can yield less irregular structural distortions on the interface and more consistent orientations of DMI vectors, which are more beneficial for cycloid spin textures.

One may ask whether the charge redistribution $\Delta\rho(\vec{r})$ is dominantly contributed by the interfacial transfer of the free electrons from metal substrates or by the weak bonds on the interface. We constructed a fictitious heterostructure system that combines Cu(111) substrate and freestanding monolayer $CrI_3$ without structural relaxations (i.e., no distortions caused by interlayer bonds). The interlayer distance of the fictitious system is kept the same as the fully relaxed structure. The $\Delta\rho(\vec{r})$ of such a fictitious system is shown in Supplemental Material (Fig. S2) [51], which also displays a prominent charge redistribution within $CrI_3$ even in the absence of structural distortion induced by Cu substrate. Without structural distortion, the $\Delta\rho(\vec{r})$ of the fictitious system better preserves the $C_3$ symmetry of the free-standing $CrI_3$. To further quantify the magnitude of substrate-induced charge redistribution, we calculated the integration over the absolute value of charge redistribution $\Delta\rho^{Total} = \int |\Delta\rho(\vec{r})| d^3r$, where the integration range is the whole supercell. The $\Delta\rho^{Total}$ = 6.78 a.u. (i.e., 6.78 electrons) for $CrI_3$/Cu(111), which is about 30 times higher than $\Delta\rho^{Total}$ = 0.24 a.u. in $CrI_3$/BL h-BN. On the other hand, $\Delta\rho^{Total}$ of the fictitious $CrI_3$/Cu(111) system is 5.58 a.u., which is slightly smaller than that of $CrI_3$/Cu(111) with fully optimized atomic coordinates. These results confirm the transfer of free electrons plays the dominant role in causing the imbalance of charge density in top and bottom layers of iodine ions and inducing a strong DMI, while the weak bonds mainly cause the fluctuations of magnetic exchange parameters.

Another important parameter for a magnetic material is the single-ion magnetic anisotropy energy (MAE), which is crucial to stabilize the long-range magnetic ordering against thermal fluctuations in 2D materials. The MAE of five heterostructures are summarized in Supplemental Material (Table S7) [51]. We also calculate the MAE of monolayer $CrI_3$ and $CrCl_3$ with 0.774 meV/Cr and 0.030 meV/Cr, respectively, which agree well with previous study which obtained 0.803 meV/Cr and 0.025 meV/Cr for $CrI_3$ and $CrCl_3$, respectively [67]. Bilayer h-BN and graphene only slightly change the MAE of monolayer $CrI_3$, which obtained the MAE value of 0.695 meV/Cr and 0.720 meV/Cr, respectively. In contrast, Au and Cu substrates prominently affect the MAE of $CrI_3$ and $CrCl_3$. Notably, the MAE of $CrI_3$ also sensitively depends on the stacking order between $CrI_3$ and Au (or Cu) substrates. We compute the MAE for different stacking modes, as shown in Fig. 4. Whereas the fluctuations of MAE in $CrI_3$/hBN and $CrCl_3$/Au(111) are negligible, the MAE of $CrI_3$/Au(111) shows drastic switches from in-plane to out-of-plane magnetic anisotropy as $CrI_3$ shifts relative to the Au(111) surface. Again, the fluctuation of MAE is due to the weak bonds on the interface, which can change with the stacking order. When the Au substrate is present, the MAE of monolayer $CrCl_3$ even turns negative value, which corresponds to an in-plane magnetic anisotropy. The in-plane magnetic anisotropy of $CrCl_3$/Au(111) is not affected by the stacking arrangement.

We remark that the lattice mismatch between $CrX_3$ and substrates creates Moiré patterns which can lead to modulation of both electronic and magnetic properties of $CrX_3$. Since the MAE and magnetic exchange interactions are sensitive to the stacking order between $CrX_3$ and metal substrate, the Moiré potential likely leads to rich non-collinear magnetic ordering in a longer range. The dependence of MAE on stacking order suggests there are regions with in-plane magnetism and those favoring out-of-plane magnetism on the superlattice. Furthermore, while Cu(111) surface has been confirmed to be free of surface reconstruction [68], the Au(111) surface is known to have a complicated $(22 \times \sqrt{3})$ surface reconstruction [69]. Such reconstruction can also affect the magnetic properties of $CrX_3$ in realistic conditions and calls for further examination.

## CONCLUSIONS

In summary, we have compared the magnetic properties of $CrX_3$ on diverse types of substrates using first-principles calculations. Although the AB-stacking BL h-BN or graphene has neglectable impact on the magnetic properties of $CrX_3$, metal substrates such as gold and copper induce strong DMI between neighboring Cr atoms. The huge

DMI in $CrI_3$/Cu(111) and $CrI_3$/Au(111) can be well explained by the strong spin-orbit interaction from iodine atoms combined with the broken spatial inversion symmetry caused by the interfacial charge redistribution and weak interlayer bonds. Interlayer interactions can also cause sizable modulation of magnetic exchange interactions and single-ion MAE in magnetic monolayers. Our research demonstrates the crucial effects of substrates in tuning the magnetic properties of 2D magnetic semiconductors like $CrX_3$ (X = Cl, I) and proposes a new class of practical material systems for realizing strong DMI. And heterostructures of vdW magnets and metal substrates are potential candidates for hosting chiral spin textures or other non-collinear magnetic orderings.

## ACKNOWLEDGMENTS

This work was supported by the National Natural Science Foundation of China (12104080, 91961204) and XingLiaoYingCai Project of Liaoning province, China (XLYC1905014). The authors acknowledge the computer resources provided by the Supercomputing Center of Dalian University of Technology and Shanghai Supercomputer Center.

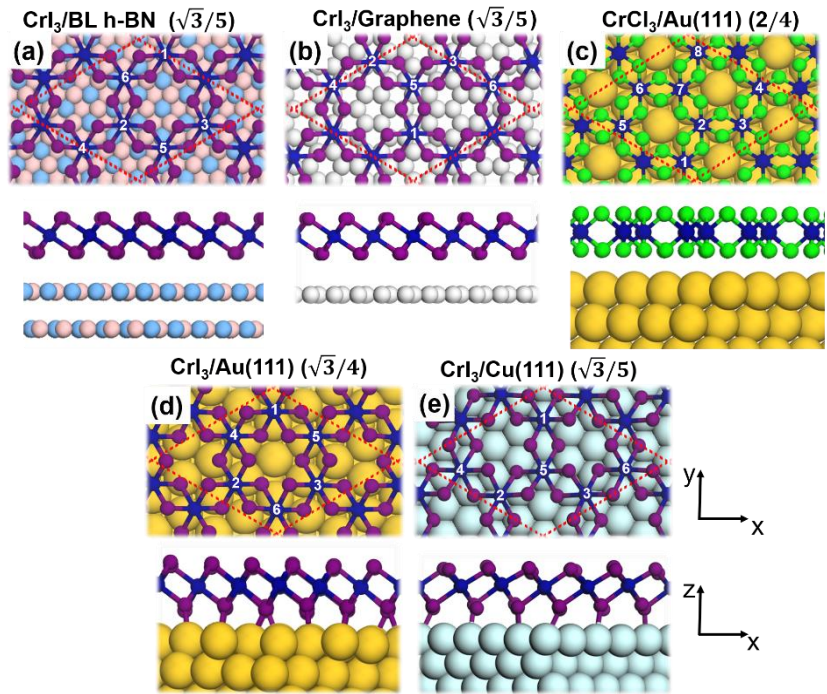

**FIG. 1.** The top and side views of (a) CrI$_3$/BL h-BN, (b) CrI$_3$/graphene, (c) CrCl$_3$/Au(111), (d) CrI$_3$/Au(111) and (e) CrI$_3$/Cu(111). Here CrI$_3$/BL h-BN ($\sqrt{3}$/5) represents a $\sqrt{3} \times \sqrt{3}$ supercell of CrI$_3$ combined with the 5 × 5 supercell of bilayer h-BN.

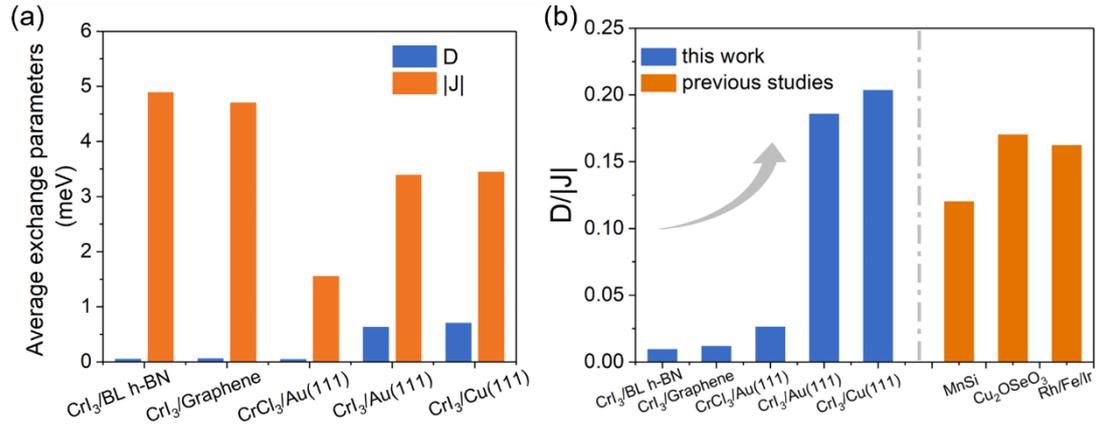

**FIG. 2.** Average magnetic exchange parameters of five CrX$_3$-substrate systems. (a) The averaged exchange parameters D and |J| between nearest-neighbor Cr pairs; (b) the ratios $\frac{D}{|J|}$ of present in comparison with and the other computational studies [59–61].

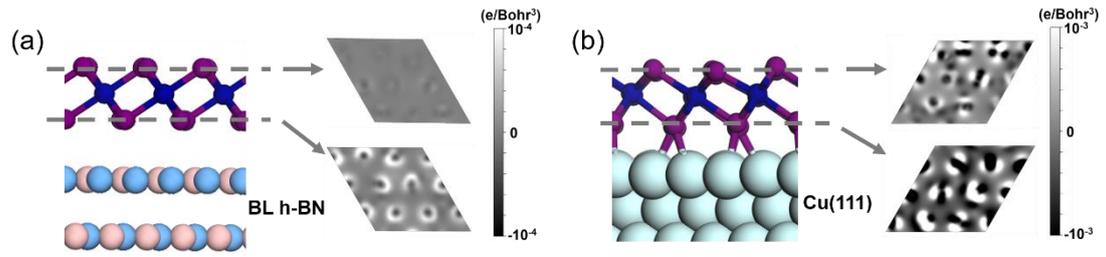

**FIG. 3.** The charge density redistribution $\Delta\rho(\vec{r})$ caused by substrates within the plane of top-layer and bottom-layer iodine atoms in (a) CrI$_3$/BL h-BN and (b) CrI$_3$/Cu(111).

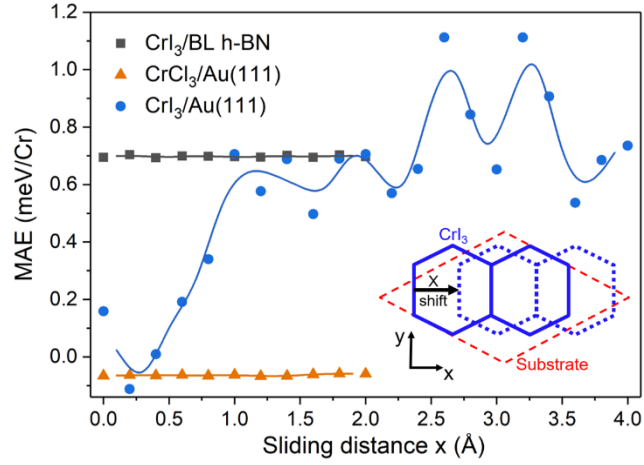

**FIG. 4.** Magnetic anisotropy energy (MAE) as a function of sliding distance of $CrX_3$ in the x direction of $CrI_3$/BL h-BN, $CrCl_3$/Au(111), and $CrI_3$/Au(111). The origin of the horizontal axis corresponds to the optimized structure shown in Fig. 1(a). $CrI_3$ shifts along the x direction by 0.2 Å per step relative to substrate. The curves are moving averages of the data points.

**Table 1.** The lattice mismatch (δ), interlayer distance (d) and binding energy ($E_b$) for five magnetic heterostructures.

|  | $CrI_3$/BL h-BN | $CrI_3$/Gr | $CrCl_3$/Au | $CrI_3$/Au | $CrI_3$/Cu |
|---|---|---|---|---|---|
| **δ (%)** | 1.72 | 0.826 | 1.41 | 1.40 | 2.87 |
| **d (Å)** | 3.60 | 3.71 | 3.14 | 2.86 | 2.47 |
| **$E_b$ (meV/Å$^2$)** | -5.03 | -13.43 | -30.67 | -50.34 | -78.88 |

# Supplemental Material:
# Strong Dzyaloshinskii-Moriya Interaction in Monolayer CrI$_3$ on Metal Substrates

Fan Zhang[1], Xueao Li[1], Yabei Wu[2,3], Xiaolong Wang[1], Jijun Zhao[1]*, Weiwei Gao[1]*

1. Key Laboratory of Material Modification by Laser, Ion and Electron Beams (Dalian University of Technology), Ministry of Education, Dalian 116024, China

2. Department of Materials Science and Engineering and Shenzhen Institute for Quantum Science & Engineering, Southern University of Science and Technology, Shenzhen, Guangdong, 518055, China

3. Guangdong Provincial Key Lab for Computational Science and Materials Design, and Shenzhen Municipal Key Lab for Advanced Quantum Materials and Devices, Southern University of Science and Technology, Shenzhen, Guangdong 518055, China

* Corresponding Authors. Email: weiweigao@dlut.edu.cn, zhaojj@dlut.edu.cn

We plot the the magnitude of displacements of CrI$_3$ deposited on Au(111) and Cu(111) substrates, respectively. The results are shown in FIG. S1. The arrows represent the direction and distance in which each atom moves. Note that the length of the arrow magnifies the atom shifting distance by a factor of 20.

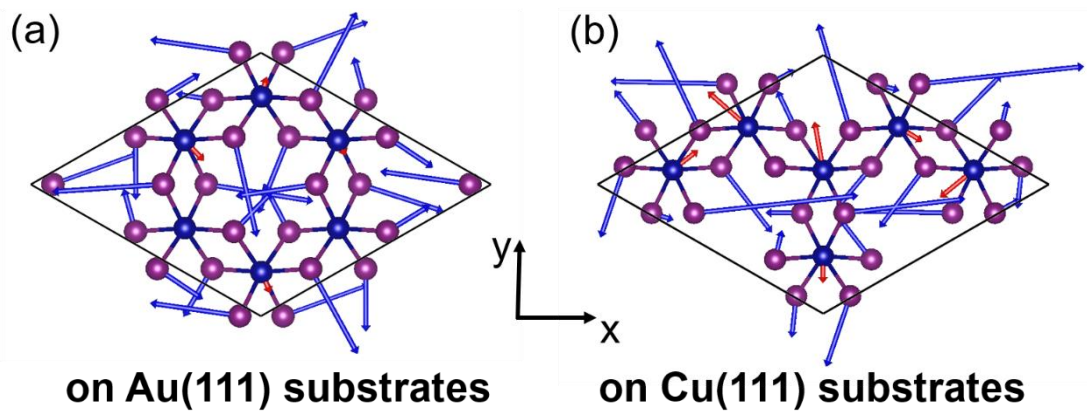

**FIG. S1.** The magnitude of displacements of Cr and I atoms in x-y plane deposited on Au(111) and Cu(111) substrates.

In order to investigate the charge redistribution is derived from the interfacial transfer or weak bonds. Here, we constructed a fictitious system that combines the free-standing Cu(111) substrate and free-standing monolayer $CrI_3$ without structural relaxations (i.e., no distortions caused by interlayer bonds), which is different from the structure of $CrI_3$/Cu(111) with fully optimized atomic coordinates in Figure 3b. The charge density redistribution $\Delta\rho(\vec{r})$ of such fictitious system is shown in FIG. S2.

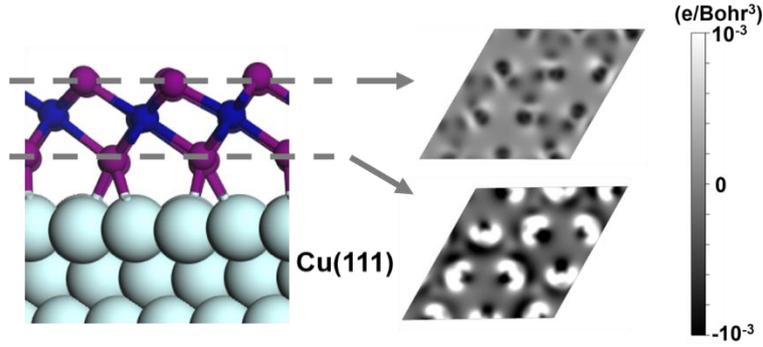

**FIG. S2.** The charge density difference with 2D slices of the top and bottom iodine atoms layers for $CrI_3$/Cu(111) without structural relaxations.

From the charge density redistribution $\Delta\rho(\vec{r})$ in FIG. S2, we note a significant difference between charge density of top and bottom layer of iodine ions, and then leading a prominent charge redistribution within $CrI_3$ even in the absence of structural distortion induced by Cu substrate, which is on the same order $10^{-3}$ e/Bohr$^3$, similar to the situation shown in Figure 3b of the main text. To further quantify the difference, the $\Delta\rho^{Total} = \int |\Delta\rho(r)|d^3r$ is calculated. The calculated $\Delta\rho^{Total}$ is 5.58 a.u. in this fictitious system $CrI_3$/Cu(111), which is slightly smaller than that of optimized $CrI_3$/Cu with 6.78 a.u.

In order to investigate the effect of the number of atomic layers in the substrates, we further modelled $CrCl_3$/Au(111) and $CrI_3$/Cu(111) heterostructures with four atomic layers as substrates. The magnetic parameters are summarized in Table S1.

The Heisenberg exchange interaction $J$ is -1.96 meV in $CrCl_3$/Au(111), which is close to -2.08 meV in three atomic layers as shown in Table S4. While $D$ is 0.0383 meV, leading to smaller $\frac{D}{|J|}$ of 0.0195 comparable to 0.0352 in three atomic layers. In addition, although the exchange parameters $J$ and $D$ in $CrI_3$/Cu(111) with four atomic layers are both much larger than that with three atomic layers in Table S6, they also lead to a large $\frac{D}{|J|}$ and favors to the practical creation of real-space topological spin textures. This indicates that the substrates with four atomic layers do not change our conclusions.

**Table S1.** Magnetic parameters of $CrCl_3$/Au(111) and $CrI_3$/Cu(111) with four atomic layers as substrates. 2-7 represents the pair between Cr2 and Cr7. Note that S = 3/2 is used when extracting the magnetic parameters. The units of energy unit are meV.

|  | $CrCl_3$/Au(111) 2-7 | $CrI_3$/Cu(111) 1-5 |
|---|---|---|
| **$J$ (meV)** | -1.96 | -9.36 |
| **($D^x$, $D^y$, $D^z$) (meV)** | (-0.0207, -0.0012, -0.0322) | (-0.734, -0.256, -0.800) |
| **$D$ (meV)** | 0.0383 | 1.115 |
| **$D/|J|$** | 0.0195 | 0.119 |

The exchange parameters of the nearest-neighbor Cr pairs within the supercells of five heterostructures and the results are summarized in Table S2-S6. Note that three nearest-neighbor Cr pairs are selected in $CrI_3$/bilayer-hBN and $CrI_3$/Graphene, and all the nearest-neighbor Cr pairs within of $CrCl_3$/Au(111), $CrI_3$/Au(111) and $CrI_3$/Cu(111) are calculated.

**Table S2.** Magnetic parameters of $CrI_3$/bilayer-hBN with nearest-neighbor Cr pairs. 2-6 represents the pair between Cr2 and Cr6. Note that S = 3/2 is used when extracting the magnetic parameters. The units of energy unit are meV.

|  | 2-6 | 2-5 | 3-5 |
|---|---|---|---|
| $J$ (meV) | -4.82 | -4.79 | -5.05 |
| $(D^x, D^y, D^z)$ (meV) | (-0.0359, -0.00164, -0.0374) | (0.0178, 0.0336, -0.0206) | (0.0115, -0.0262, -0.0306) |
| $D$ (meV) | 0.0519 | 0.0432 | 0.0419 |
| $D/|J|$ | 0.0108 | 0.00902 | 0.00830 |

**Table S3.** Magnetic parameters of CrI$_3$/Graphene with nearest-neighbor Cr pairs. 1-5 represents the pair between Cr1 and Cr5. Note that S = 3/2 is used when extracting the magnetic parameters. The units of energy unit are meV.

|  | 1-5 | 2-5 | 2-4 |
|---|---|---|---|
| **J (meV)** | -4.72 | -4.77 | -4.62 |
| **($D^x$, $D^y$, $D^z$) (meV)** | (0.0614, -0.00313, -0.0105) | (0.00716, 0.000690, 0.0391) | (-0.0364, 0.0509, -0.0121) |
| **D (meV)** | 0.0624 | 0.0397 | 0.0638 |
| **D/\|J\|** | 0.0132 | 0.00832 | 0.0138 |

**Table S4.** Magnetic parameters of CrCl$_3$/Au(111) with every nearest-neighbor Cr pairs. 2-7 represents the pair between Cr2 and Cr7. Note that S = 3/2 is used when extracting the magnetic parameters. The units of energy unit are meV.

|  | 2-7 | 2-1 | 2-3 | 6-3 | 6-5 | 6-7 |
|---|---|---|---|---|---|---|
| **J (meV)** | -2.08 | -0.502 | 2.07 | -2.08 | -0.503 | 2.07 |
| **($D^x$, $D^y$, $D^z$) (meV)** | (-0.0108, 0.00452, -0.0723) | (0.0000528, 0.0207, -0.0271) | (-0.0217, -0.0251, -0.0348) | (0.0159, 0.00339, -0.0242) | (-0.00191, 0.00946, -0.0266) | (-0.0216, -0.0247, -0.0346) |
| **D (meV)** | 0.0733 | 0.0341 | 0.0481 | 0.0292 | 0.0283 | 0.0477 |
| **D/\|J\|** | 0.0352 | 0.0679 | 0.0232 | 0.0140 | 0.0563 | 0.0230 |
|  | 3-4 | 1-4 | 4-5 | 7-8 | 5-8 | 1-8 |
| **J (meV)** | -0.502 | 2.07 | -2.08 | -0.496 | 2.07 | -2.08 |
| **($D^x$, $D^y$, $D^z$) (meV)** | (-0.00489, -0.0131, 0.0263) | (0.0501, 0.0256, 0.00850) | (-0.00586, 0.00427, -0.00136) | (0.00473, -0.0119, 0.0296) | (0.0220, 0.0261, 0.0349) | (0.0340, -0.0245, 0.0265) |
| **D (meV)** | 0.0299 | 0.0569 | 0.00738 | 0.0322 | 0.0489 | 0.0496 |
| **D/\|J\|** | 0.0596 | 0.0275 | 0.00355 | 0.0649 | 0.0236 | 0.0238 |

**Table S5.** Magnetic parameters of CrI$_3$/Au(111) with every nearest-neighbor Cr pairs. 1-4 represents the pair between Cr1 and Cr4. Note that S = 3/2 is used when extracting the magnetic parameters. The units of energy unit are meV.

|  | 1-4 | 1-5 | 1-6 | 2-4 | 2-5 | 2-6 | 3-4 | 3-5 | 3-6 |
|---|---|---|---|---|---|---|---|---|---|
| **J (meV)** | 1.28 | -2.79 | -3.49 | -8.12 | -4.39 | -2.25 | -2.34 | -3.66 | 2.15 |
| **($D^x$, $D^y$, $D^z$) (meV)** | (0.258, -0.825, 0.448) | (-0.057, -0.254, -0.565) | (-0.466, -0.326, 0.098) | (-0.159, 0.122, -0.388) | (0.180, -0.103, 0.513) | (0.221, -0.393, -0.287) | (0.489, -0.279, 0.313) | (-0.464, -0.231, -0.264) | (-0.230, 0.460, -0.524) |
| **D (meV)** | 0.974 | 0.622 | 0.577 | 0.437 | 0.554 | 0.534 | 0.644 | 0.581 | 0.734 |
| **D/\|J\|** | 0.761 | 0.223 | 0.165 | 0.0538 | 0.126 | 0.237 | 0.275 | 0.159 | 0.341 |

**Table S6.** Magnetic parameters of $CrI_3$/Cu(111) with every nearest-neighbor Cr pairs. 1-4 represents the pair between Cr1 and Cr4. Note that S = 3/2 is used when extracting the magnetic parameters. The units of energy unit are meV.

|  | 1-4 | 1-5 | 1-6 | 2-4 | 2-5 | 2-6 | 3-4 | 3-5 | 3-6 |
|---|---|---|---|---|---|---|---|---|---|
| *J* (meV) | -2.57 | -7.25 | -5.21 | -1.60 | -4.20 | 0.0543 | -5.72 | -2.57 | -1.82 |
| ($D^x$, $D^y$, $D^z$) (meV) | (-0.284, 0.203, 0.131) | (-0.503, -0.0470, -0.694) | (0.110, 0.442, 0.425) | (0.228, 0.161, -0.509) | (0.137, 0.043, 0.155) | (0.953, -0.028, 1.122) | (-0.884, -0.010, -0.647) | (-0.299, -0.271, 0.166) | (0.279, -0.195, -0.556) |
| *D* (meV) | 0.373 | 0.858 | 0.623 | 0.581 | 0.212 | 1.472 | 1.095 | 0.436 | 0.652 |
| *D*/|*J*| | 0.145 | 0.118 | 0.120 | 0.363 | 0.0505 | 27.1 | 0.191 | 0.170 | 0.358 |

Table S7. Magnetic anisotropy energy (MAE) for five heterostructures, which is defined as $E(\|) - E(\perp)$. Note that they are calculated with the optimized structure in Figure 1.

|  | $CrI_3$ | $CrCl_3$ | $CrI_3$/BL h-BN | $CrI_3$/Gr | $CrCl_3$/Au | $CrI_3$/Au | $CrI_3$/Cu |
|---|---|---|---|---|---|---|---|
| **MAE (meV/Cr)** | 0.774 | 0.030 | 0.695 | 0.720 | -0.066 | 0.158 | 0.866 |